\documentclass[aps,amsmath,amssymb,showpacs,showkeys]{revtex4}
\usepackage[dvips]{graphicx,color}
\usepackage{times}
\usepackage{xcolor}
\usepackage[%
  colorlinks=true,
  urlcolor=blue,
  linkcolor=red,
  citecolor=blue
]{hyperref}

\begin{document}
\title{Gravitational Waves in $\mathbf{f(R)}$ Gravity Power Law Model}

\author{Dhruba Jyoti Gogoi}
\email[Email: ]{moloydhruba@yahoo.in}

\affiliation{Department of Physics, Dibrugarh University,
Dibrugarh 786004, Assam, India}

\author{Umananda Dev Goswami}
\email[Email: ]{umananda2@gmail.com}

\affiliation{Department of Physics, Dibrugarh University,
Dibrugarh 786004, Assam, India}

%\date{}
\begin{abstract}
We investigate the different polarization modes of 
Gravitational Waves (GWs) in $f(R)$ gravity power law model in de Sitter 
space. It is seen that the massive scalar field polarization mode exists in 
this model. The mass of the scalar field depends highly on the background 
curvature and the power term $n$. However, we found that the model doesnot 
exhibit a massive scalar mode for $n=2$ and instead it shows a breathing mode 
in addition to the tensor plus and cross modes. Thus mass of the scalar field 
is found to vary with $n$ within the range $1 \leq n \leq 2$.
\end{abstract}


\maketitle

\section*{1. Introduction}
Inspired purely by intellectual curiosity, the first modification of 
Einstein's gravity was attempted by Weyl in 1919 \cite{weyl_neue_1919} and 
then by Eddington in 1923 \cite{eddington_mathematical_1924} with the 
introduction of higher order curvature invariants in the gravitational action. 
Due to the lack of experimental motivations, these types of modifications were 
ignored untill 1962. During 1962 for the first time it was realized that 
modification of gravitational action could have some merits. In fact, the 
gravity from General Relativity (GR) is not renormalizable and thus it is not 
possible to quantize it according to conventional methods. In 1962, Utiyama 
and DeWitt showed that the renormalization of gravity at one loop is possible 
if one modifies the Einstein-Hilbert action by higher order curvature terms
\cite{utiyama_renormalization_1962}. Later, a number of drawbacks of GR
have been observed including its inability to explain cosmic acceleration.
To overcome these drawbacks of GR several new theories have been proposed
suggesting further modifications of GR. Among them, $f(R)$ gravity is a
prominent one which keeps itself in stand against the challenges and problems
it faced till now \cite{sotiriou_fr_2010, nojiri_2011, nojiri_2007}. One of the
important features of $f(R)$ gravity is its less complexity compared to other
modified theories of gravity \cite{nojiri_2017}. 

After around 100 years of Einstein's prediction of existence of Gravitational 
Waves (GWs), the Laser Interferometer Gravitational Wave Observatory (LIGO) 
Scientific Collaboration announced the detection of GWs for the 
first time on September 14, 2015. This first GW event was named as GW150914 
\cite{abbott_observation_2016}. This observation was later followed by many 
other events including GW151226 \cite{abbott_gw151226_2016}, GW170104 
\cite{abbott_gw170104_2017}, GW170814 \cite{abbott_gw170814:_2017} and 
GW170817 \cite{abbott_gw170817:_2017}. These events opened new 
directions in testing GR and 
modified theories of gravity. Therefore the study of GWs in modified gravity 
will play a very important role in modifying GR and on the predictions of GR. 
The experimental results obtained from the sector of GWs can effectively 
constrain 
the modified gravity models to a fair extent and can even contribute to the 
bottom-up approach in forming or proposing modified gravity theories and 
models in an efficient way.

In the metric formalism of $f(R)$ gravity, the GWs have other modes of 
polarizations besides the tensor modes of polarization found in GR. 
Generation and possibilities of detection of GWs in $f(R)$ gravity were 
studied by C. Corda using the Starobinsky model in \cite{corda01} and using
a model of the form $ f(R) = R+R^{-1}$ in \cite{corda02}.
These studies indicate the existence of massive mode of GW radiation \cite{corda03} in $f(R)$ 
theory in metric formalism. In a recent study, the propagating degrees of freedom of GWs 
in $f(R)$ gravity are found to be 3 \cite{myung_propagating_2016}, and in 
another recent study on the Starobinsky model \cite{liang_polarizations_2017} 
showed that there exists a mixed state of massless breathing mode and massive 
scalar mode of polarization besides the usual tensor modes in the model. Thus 
the Newman-Penrose (NP) formalism cannot be applied 
in that study due to the existence of massive scalar modes of GW polarization. 
In our study, we'll restrict ourselves to the power law model of $f(R)$ gravity 
given by $f(R) = \alpha R^n$, where $\alpha$ is an arbitrary constant and $n$ 
is a real number. This model has been pursued in \cite{jaime_about_2013} to study 
the late time acceleration in 
the Universe. It is already established that realistic $f(R)$ gravity models may unify inflation, radiation dominance, matter dominance and dark energy \cite{Nojirim1, Nojirim2, Nojirim3, Nojirim4, Nojirim5}. 
In the Refs. \cite{capozziello_quintessence_2003, carloni_cosmological_2005, 
goswami_cosmological_2013} the power law model was discussed for the different special 
cases along with $n=-1$ and $n=3/2$, in which $n=3/2$ case is conformally 
equivalent to Liouville field theory. In another paper \cite{faraoni_r_2011}, 
the model was used for invoking the chameleon mechanism to work within the solar 
system and in which it was shown that for this to happen the value of $n$ is 
required to be very close to $1$. The result rendered from the work implied 
that the model is a poor candidate for a realistic alternative to dark energy.
The cosmological dynamics of the model also has been studied in Ref. \cite{carloni_cosmological_2005}. It is to be noted that although
the model used in our case is already ruled out in solar system tests \cite{zakharov}, 
we have noticed that in several studies \cite{faraoni_r_2011,capozz, borka, zakharov2} 
the model has been used as a toy model. These studies have motivated 
us to use this model for its simple form and 
more importantly because of the existence of massless scalar mode of 
polarization of GWs even in de sitter space. Also we think that a wide range of 
studies with this sort of new result will definitely help to decisively rule 
out or accept the model from future astrophysical/cosmological observations.

In this work, we'd like to test the model in the GW regime to find out the 
possibility of its viability and the drawbacks. Another important point of this 
model is that it is the only model in $f(R)$ theories in the metric formalism which can give pure massless breathing mode as a special case. Here we'll study the 
model with the special case of $n=2$ along with the stability and polarization 
modes.
 
The paper is organized as follows. In the next section the variation of 
mass of the scalar field in de Sitter space with the power term $n$ is studied.
In the third section the stability of the model is studied in de Sitter space.
%In the fourth section, the contents of the theory and the model are studied with 
%the help of operator formalism. In there, we tried to show the contents of the 
%model for a specific case of $n=2$ and we show that in $f(R)$ theory of 
%modified gravity a pure massless scalar field is also possible as in case of 
%Scalar Tensor theory. 
The fourth section contains a study towards the 
equivalence of the theory with the Scalar-Tensor theory and the required 
conditions for such equivalence. The tensor and the scalar polarization modes 
are studied with the help of geodesic deviation in the fifth section and with 
the help of Newman-Penrose formalism in the sixth section. In section 
seven, we have discussed a possible way to check the validity of the model 
experimentally. In the last section, we conclude the paper with a very brief 
discussion of the results and the future aspects of the model in such type of 
studies.

\section*{2. Scalar and tensor fields from the model} \label{sec2}
The action of $f(R)$ gravity is given by
\begin{equation}
S =\dfrac{1}{2\kappa} \int d^4x \sqrt{-g}\, f(R),
\label{eq1}
\end{equation} 
where $f(R)$ is a function of Ricci curvature $R$. The vacuum field equations 
obtained from this action are given by
\begin{equation}\label{eq2}
 f'(R)R_{\mu\nu} -\dfrac{1}{2} f(R)g_{\mu\nu} - \nabla_\mu \nabla_\nu f'(R) 
 + g_{\mu\nu}\, \square f'(R) = 0, 
\end{equation} 
where $\square = g^{\mu\nu}\,\nabla_\mu \nabla_\nu$ and the prime over $f(R)$ 
denotes the derivative with respect to $R$. Taking trace of vacuum 
field Eq. (\ref{eq2}), we obtain
\begin{equation} \label{eq3}
f'(R) R +3\, \square f'(R) - 2 f(R) = 0.
\end{equation}

Now, we assume that there is a propagating GWs in the spacetime, which will 
perturb the metric around its background value. Since the perturbation is 
usually very small and hence if we consider the background metric as 
$\bar{g}_{\mu\nu}$ then to the first order of perturbation value $h_{\mu\nu}$, 
we may express the spacetime metric as
\begin{equation}
g_{\mu\nu}=\bar{g}_{\mu\nu} + h_{\mu\nu},\;\; \mbox{where}\; \vert h_{\mu\nu} \vert << \vert \bar{g}_{\mu\nu} \vert.
\label{eq4}
\end{equation}
In view of this perturbation, expanding the Ricci tensor and the Ricci scalar 
upto the first order of $h_{\mu\nu}$, we may write
\begin{align}
\notag
 R_{\mu\nu}& \simeq \bar{R}_{\mu\nu} + \delta R_{\mu\nu} + O(h^2)\\  \notag
&= \bar{R}_{\mu\nu} - \dfrac{1}{2} (\nabla_\mu \nabla_\nu 
h - \nabla_\mu \nabla^{\lambda} h_{\lambda\nu}- \nabla_\nu \nabla^{\lambda} h_{\mu\lambda} \\
& + \square h_{\mu\nu}) + O(h^2)
\label{eq5}
\end{align}
and 
\begin{align}
\notag
 R &\simeq \bar{R} + \delta R + O(h^2)\\
 & =\bar{R} - \square h + \nabla^\mu \nabla^\nu h_{\mu\nu} -\bar{R}_{\mu\nu} h^{\mu\nu} + O(h^2),
\label{eq6}
\end{align}
where $\bar{R}$ is the de Sitter curvature. Thus, due to this 
perturbation the trace Eq. (\ref{eq3}) can be rewritten as
\begin{equation}
3 f''(\bar{R}) \square \delta R + \left[f''(\bar{R})\bar{R} - f'(\bar{R})\right] \delta R = 0.
\label{eq7}
\end{equation}
Using our model 
\begin{equation}
f(R)=\alpha R^n %, $f'(R)=\alpha n R^{n-1} $, and $f''(R)=\alpha n(n-1) 
%R^{n-2}$ 
\label{eq8}
\end{equation}
the Eq. (\ref{eq7}) can be expressed in the following compact form:
\begin{equation} \label{eq9}
 (\square - m^2) \delta R = 0,
\end{equation}
where
\begin{equation} \label{eq10} 
m^2 = \dfrac{(2-n)}{3(n-1)} \bar{R}.
\end{equation} 
The Eq. (\ref{eq9}) is the conventional form of the scalar field 
equation (Klein-Gordon) with $\delta R$ as the scalar field and hence here, 
$m$ can be identified as the mass of this scalar field. Moreover, since this 
scalar field corresponding to $\delta R$ is due to ripple in spacetime, so the
field can be considered as the quantized massive gravitational field.   

\begin{figure}[h]
\centerline{
\includegraphics[scale=0.4]{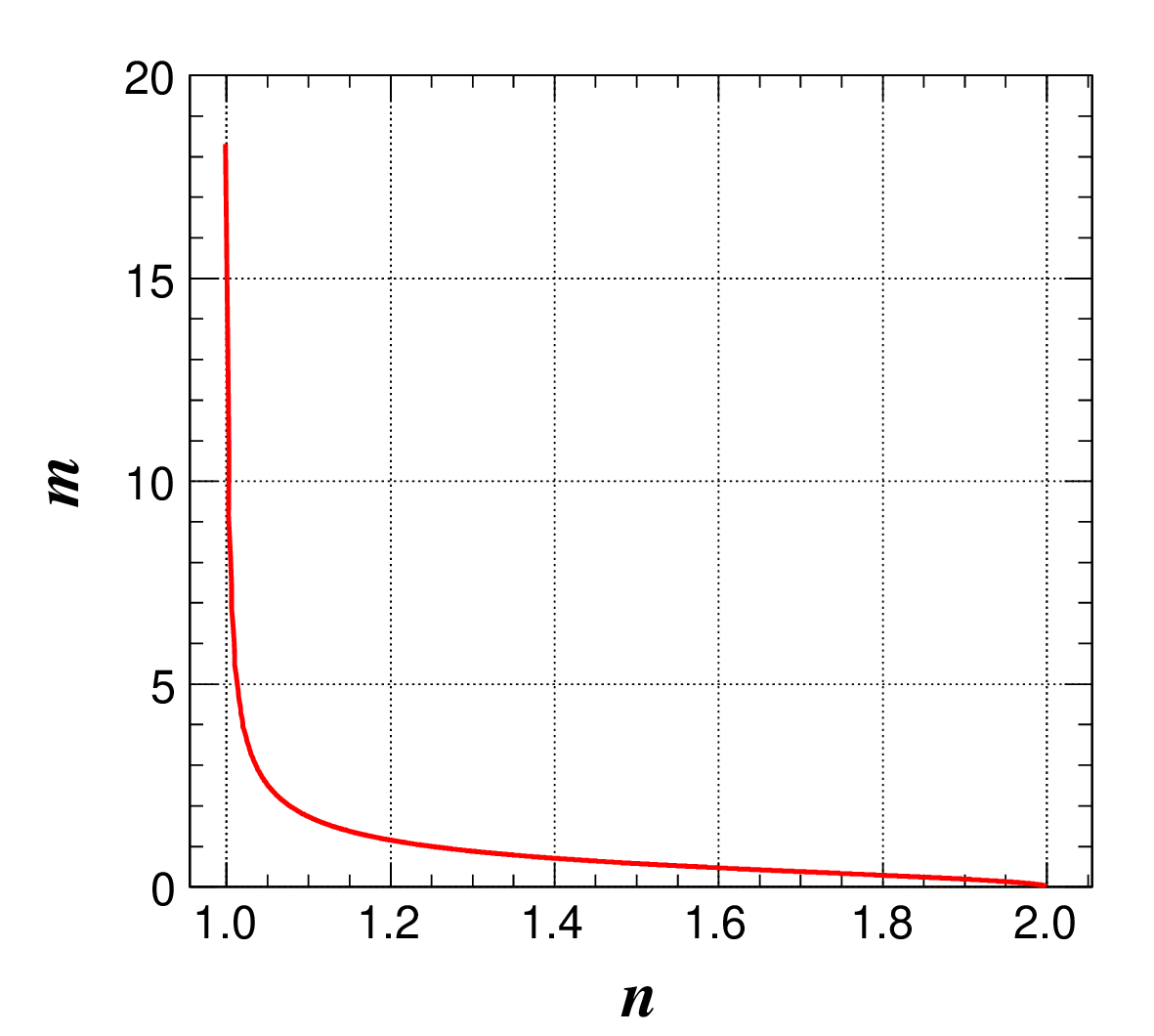}}
\caption{Variation of mass $m$ of the scalar field with respect to $n$ for the 
unit value of the background curvature $\bar{R}$.}
\label{fig1}
\end{figure}

From the expression (\ref{eq10}) we see that the mass term of the scalar field 
depends highly on the exponent term $n$ and the background curvature $\bar{R}$.
This indicates that near the massive objects like neutron stars etc., where
the background curvatures are very large, the mass of the scalar form of 
gravitational field will increase and consequently this will slow down the 
propagation speed of the massive mode of GWs in such regions according to this 
particular model. Further, for a massive scalar field or massive mode of GWs $n$
should lie in between $1$ and $2$ beyond which it will result 
tachyonic instabilities. When $n=1$ the model will give the Einstein's case. When 
$n$ starts to increase from $1$ to $2$ mass of the field decreases 
monotonically and finally, for $n=2$ the massive scalar mode of polarization 
will vanish. i.e. for the value $n=2$ model will give a massless scalar field. The 
variation of the scalar field mass with respect to $n$ for the unit value of 
the background curvature is shown in the Fig.\ref{fig1}. One interesting 
observation from the mass term is that when $n$ approaches to $1$ the mass of 
the scalar field increases very rapidly and for $n=1$ the mass becomes 
infinite. Thus for $n=1$ the scalar mode vanishes and only the massless 
spin-2 modes propagate. 

The tensor field Eq. (\ref{eq2}) in vacuum for the power law model 
(\ref{eq8}) takes the from: 
\begin{equation}
 n R^{n-1} R_{\mu\nu} -\dfrac{1}{2} g_{\mu\nu}  R^n - n \nabla_\mu \nabla_\nu R^{n-1} + n\,g_{\mu\nu} \square R^{n-1}  = 0.
\label{eq11}
\end{equation} 
Perturbing this field equation around the de Sitter curvature $\bar{R}$ (see 
the Eq. (\ref{eq5})) we may write,
%\begin{eqnarray}
%\nonumber
%n\bar{R}\bar{R}_{\mu\nu} +n(n-1)\bar{R}_{\mu\nu} \delta R -\dfrac{1}{2} \bar{g}
%  _{\mu\nu} \bar{R}^2 +n\bar{R}\,\delta R_{\mu\nu}-\\ 
%  \dfrac{1}{2} n\, \bar{g}_{\mu\nu}\bar{R}\,\delta R
%-\,n(n-1)\nabla_\mu \nabla_\nu(\delta R)+n(n-1)\bar{g}_{\mu\nu}\square 
%  \delta R = 0.
%\label{eq12}
%\end{eqnarray}

%%%
\begin{align}
\label{eq12}
\notag
n\bar{R}\bar{R}_{\mu\nu} +n(n-1)\bar{R}_{\mu\nu} \delta R -\dfrac{1}{2} \bar{g}
  _{\mu\nu} \bar{R}^2 - \dfrac{1}{2} n\, \bar{g}_{\mu\nu}\bar{R}\,\delta R \\ 
+n\bar{R}\,\delta R_{\mu\nu}-\,n(n-1)\nabla_\mu \nabla_\nu(\delta R)+n(n-1)\bar{g}_{\mu\nu}\square 
  \delta R = 0.
\end{align}
%%%

In the ideal case, i.e. when $R=\bar{R}$, the de Sitter curvature should 
satisfy the above field equation. Thus for the ideal case the above equation 
gives,
  \begin{equation} \label{eq13}
  n\bar{R}_{\mu\nu}=\dfrac{1}{2}\bar{g}_{\mu\nu}\bar{R}.
  \end{equation}
Trace of this equation gives
$$ n \bar{R} = 2 \bar{R},$$
which can be written as
\begin{equation} \label{eq14}
n\bar{R}^2 - 2\bar{R}^2=0.
\end{equation}
Again the trace of Eq.\ (\ref{eq12}) is,
$$ n\bar{R}^2 - 2\bar{R}^2 + n(n-2)\bar{R}\delta R + 3n(n-1)\square \delta R = 0.$$
Using Eq.\ (\ref{eq14}) in the above equation we find,
  \begin{equation}
 (\square - m^2) \delta R = 0, 
   \end{equation}
 where
 $m^2 = \dfrac{(2-n)}{3(n-1)} \bar{R}.$ This equation is identical to the Eq.\ (\ref{eq9}). For $n=2$ above equation reduces to,
  \begin{equation}\label{eq15}
  \square\delta R = 0.
  \end{equation}
This is the equation of massless scalar field for the case of $n=2$. This is an interesting result, which shows that if the 
de Sitter curvature has to satisfy the field Eq. (\ref{eq12}) then the 
model (\ref{eq8}) has to take the value $n=2$, consequently, the massive scalar
mode of polarization vanishes and a pure massless scalar mode of polarization 
(also known as breathing mode of polarization) is obtained.
 
Thus the Eq. (\ref{eq15}) can be treated as a stability point for the 
model in de Sitter space which identically satisfies the stability condition 
discussed in the following section without modifying the Hubble constant in de 
Sitter space.

Again, since $n=2$ in the model (\ref{eq8}) (we'll refer this model with
$n=2$ as pure $R^2$ model) makes the scalar field massless and independent of
the the background curvature $\bar{R}$, so for simplicity we take this 
advantage to choose $\bar{R} = 0$ and perturb the tensor field equation for 
pure $R^2$ model (see the Eq. (\ref{eq2})) to give   
\begin{equation}\label{eq16}
R\delta R_{\mu \nu} - \dfrac{1}{4} R^2 \bar{g}_{\mu\nu} - \partial_\mu \partial_\nu R+\bar{g}_{\mu \nu}\square R = 0.
\end{equation}
Now we define a parameter,
\begin{equation}\label{eq17}
h_{\mu \nu} = \bar{h}_{\mu\nu}-\dfrac{1}{2}\bar{g}_{\mu\nu}\bar{h}-\bar{g}_{\mu\nu}\delta R,
\end{equation}
where taking the trace we find $\bar{h}$ as 
\begin{equation}
\bar{h} = -\, h - 4\delta R,
\end{equation}
and substituting $\bar{h}$ in the definition (\ref{eq17}), $\bar{h}_{\mu \nu}$
can be expressed as 
\begin{equation}
\bar{h}_{\mu \nu} = h_{\mu\nu}-\dfrac{1}{2}\bar{g}_{\mu\nu}h-\bar{g}_{\mu\nu}\delta R.
\end{equation}
Choosing transverse traceless gauge condition:
$ \partial^\mu \bar{h}_{\mu\nu}=0$, $\bar{h}=0$
and using the definition (\ref{eq17}) in Eq. (\ref{eq5}), we get
\begin{equation}\label{eq20}
\delta R_{\mu\nu} = \dfrac{1}{2} \left[-\,\square\bar{h}_{\mu\nu} + 2 \partial_\mu \partial_\nu (\delta R)
+ \bar{g}_{\mu\nu} \square \delta R  \right]. 
\end{equation}
Finally, using this Eq. (\ref{eq20}) and Eq. (\ref{eq15}) in the 
Eq. (\ref{eq16}), we may write
\begin{equation}\label{eq21}
\square \bar{h}_{\mu\nu} = 0,
\end{equation}
which is the usual equation that gives the tensor modes, i.e. plus and cross 
modes of polarization of GWs. The solution of this equation is
\begin{equation}\label{eq22}
\bar{h}_{\mu\nu} = e_{\mu\nu} \exp(iq_\mu x^\mu) + c.c.,
\end{equation}
where $\bar{g}_{\mu\nu} q^\mu q^\mu = 0$ and $q^\mu e_{\mu\nu}=0.$ Thus, this
clearly shows that the pure $R^2$ gravity model gives three polarization modes
of GWs, all massless, two of them are tensor modes while the other is the 
scalar or the pure breathing mode as discussed above.

\section*{3. Stability of the Model}

According to John D. Barrow and A. C. Ottewill \cite{barrow_stability_1983} the 
condition for the existence of de Sitter solution for a constant curvature 
$\bar{R}$ is
\begin{equation}\label{eq23}
\bar{R} f'(\bar{R}) = 2 f(\bar{R}).
\end{equation}
In the Einstein's case $\bar{R}=4\Lambda$ \cite{barrow_stability_1983}, where 
$\Lambda$ is the cosmological constant. For our model (\ref{eq8}), we get from 
the above condition that
\begin{equation}\label{eq24}
n \bar{R}^n = 2 \bar{R}^n.
\end{equation}
This condition is similar to the condition obtained in Eq. (\ref{eq13}), 
giving $n=2$ and which transforms the Eq. (\ref{eq9}) into the Eq. 
(\ref{eq15}) with mass for the scalar field $m=0$ as mentioned in the 
previous section. However, there is another possible solution for this 
Eq. (\ref{eq13}) is $\bar{R}= 0$, which will lead to Minkowski spacetime 
having zero background curvature.

In \cite{faraoni_stability_2005} it was shown that the model (\ref{eq8}) 
produce power law inflation 
$ a\propto t^{\bar{\alpha}} $, with
$$\bar{\alpha} = \dfrac{(n-1)(1-2n)}{(n-2)}$$ 
for generic values of $n\neq 0,1/2,1$ and as we have already seen that this model 
does not allow or admit de-Sitter solution for $n\neq2$. This will be a case 
only when a cosmological constant is added to the action corresponding to a 
term with $n = 0$ \cite{barrow_stability_1983, faraoni_stability_2005}. But for
any positive value of $n$ Minkowski space is a solution without any cosmological 
constant \cite{faraoni_stability_2005}. The stability condition for this model 
from \cite{faraoni_stability_2005} reads
\begin{equation}\label{25}
\bar{R}\dfrac{2-n}{n(n-1)}\geq 0.
\end{equation} 
This condition gives, $1<n\leq 2$ and $n< 0$. From Eq.\ (\ref{eq10}) it can be 
easily seen that the first stability condition i.e. $1<n\leq 2$ agrees well 
with the non-tachyonic range of the scalar field. $n> 2$ and $n<1$ in 
Eq.\ (\ref{eq10}) will make $m^2$ negative resulting tachyonic instabilities. 
However, the second stability condition obtained from Eq.\ (\ref{25}) i.e. 
$n< 0$ in 
Eq.\ (\ref{eq10}) can also result tachyonic instabilities. To avoid such 
tachyonic instabilities in our toy model we would not consider the situation
satisfied by the later condition. 
Therefore, the model is considered to be stable in the range $1<n\leq 2$ only. 
For $n=2$ it is identically 
satisfied without imposing any constraints on the Hubble parameter in de 
Sitter space and also shows that for any values of $n$ the Minkowski space is 
stable \cite{faraoni_stability_2005}. Thus our previous results are supported
by these inferences.

\section*{4. Equivalence with the Scalar Tensor Theory}
A straight forward way to study $f(R)$ gravity is to see its equivalence with 
the Scalar Tensor Theory (STT) \cite{chiba_1/r_2003}. This is reasonable 
because as like $f(R)$ theories, STTs also have two types of polarization 
modes, tensor modes and scalar mode. By introducing a scalar field $\phi \equiv R$, the general form of the action (\ref{eq1}) can be given as
\begin{equation}\label{eq30}
S=\dfrac{1}{2\kappa}\int d^4 x \sqrt{-g}\,\big[\psi(\phi)R-V(\phi)\big],
\end{equation}
where $\psi (\phi) = f'(\phi)$ and $v(\phi) = \phi f'(\phi)- f(\phi).$
Now, differentiating Eq. (\ref{eq30}) with respect to $\phi$, we find
\begin{equation}
f''(R)(R-\phi)=0.
\end{equation}
This shows that when $f''(R)\neq 0$, we must have $R=\phi$. Thus $f''(R)\neq 0$
is a very important condition required to be satisfied in order to compare 
the $f(R)$ theory with the STT. The pure $R^2$ model satisfies this 
condition. In metric formalism, the STT 
with Brans-Dicke parameter $\omega =0$ is equivalent to the $f(R)$ gravity.
 
Again the action (\ref{eq30}) can be written as
 \begin{equation}\label{31}
 S=\dfrac{1}{2\kappa}\int d^4 x \sqrt{-g}\big[f(\phi)+f'(\phi)(R-\phi)\big].
 \end{equation}
 The field equation obtained from the above action is given as
 \begin{align}
 \label{eq32} \notag
 G_{\mu\nu}=\dfrac{1}{f'(\phi)}\big[\nabla_\mu\nabla_\nu f'(\phi)-g_{\mu\nu}\,\square 
 f'(\phi) + \dfrac{1}{2}\,g_{\mu\nu}\lbrace f(\phi) \\
 -\phi f'(\phi)\rbrace\big].
 \end{align}
 Trace of the above equation is
 \begin{equation}
 \square f' = \dfrac{2}{3} f(\phi) - \dfrac{1}{3} \phi f'(\phi).
 \end{equation}
 For the pure $R^2$ model the Eq. (\ref{eq32}) gives
 \begin{equation} \label{eq34}
 G_{\mu\nu}=\phi^{\,-1}\big(\partial_\mu\partial_\nu \phi-g_{\mu\nu}\,\square \,\phi -\dfrac{1}{4} g_{\mu\nu}\,\phi^2\big).
 \end{equation}
And from the trace of this equation we get,
 \begin{equation}\label{eq35}
 \square\,\phi=0.
 \end{equation}
This equation is similar to the Eq. (\ref{eq15}) which gives a massless 
breathing mode of polarization. In Minkowski space $\square\,\phi \equiv \square\,\delta\phi$, since $\square\,\bar{\phi} = 0$ in this space, so the solution 
of this equation can be written as
 \begin{equation} \label{eq36}
 \delta\phi=\phi_0 \exp(ip_\mu x^\mu) + c.c.,
 \end{equation}
 where $\bar{g}_{\mu\nu}p^\mu p^\nu=0.$ In case of non zero background, a damping factor can appear in the solution. But the effect of this damping factor is very negligible and we can safely neglect the term \cite{damping_sol}. Hence, this equation is valid in both de Sitter spacetime and Minkowski spacetime. Now, combining the solutions for the 
massless tensor modes and the scalar mode, and considering that the GW is 
traveling along the z-axis with the speed of light $c=1$, we get the solution
for $h_{\mu\nu}$ as
 \begin{equation} \label{eq37}
 h_{\mu\nu}= \bar{h}_{\mu\nu}(t-z)-\bar{g}_{\mu\nu}\,\delta\phi(t-z).
 \end{equation}
Hence, for this solution we may take $q_{\mu} = \Omega\,(1,0,0,1)$ and
$p_\mu = \omega\,(1,0,0,1)$. It should be noted that for $n\neq 2$ and 
$n\neq 1$ the above solution will contain a mixed state of massive scalar 
mode and breathing mode besides the tensor modes of 
polarization \cite{liang_polarizations_2017}.

\section*{5. Geodesic Deviation and Polarization Modes}
\begin{figure}
\centerline{ 
\includegraphics[scale=0.32]{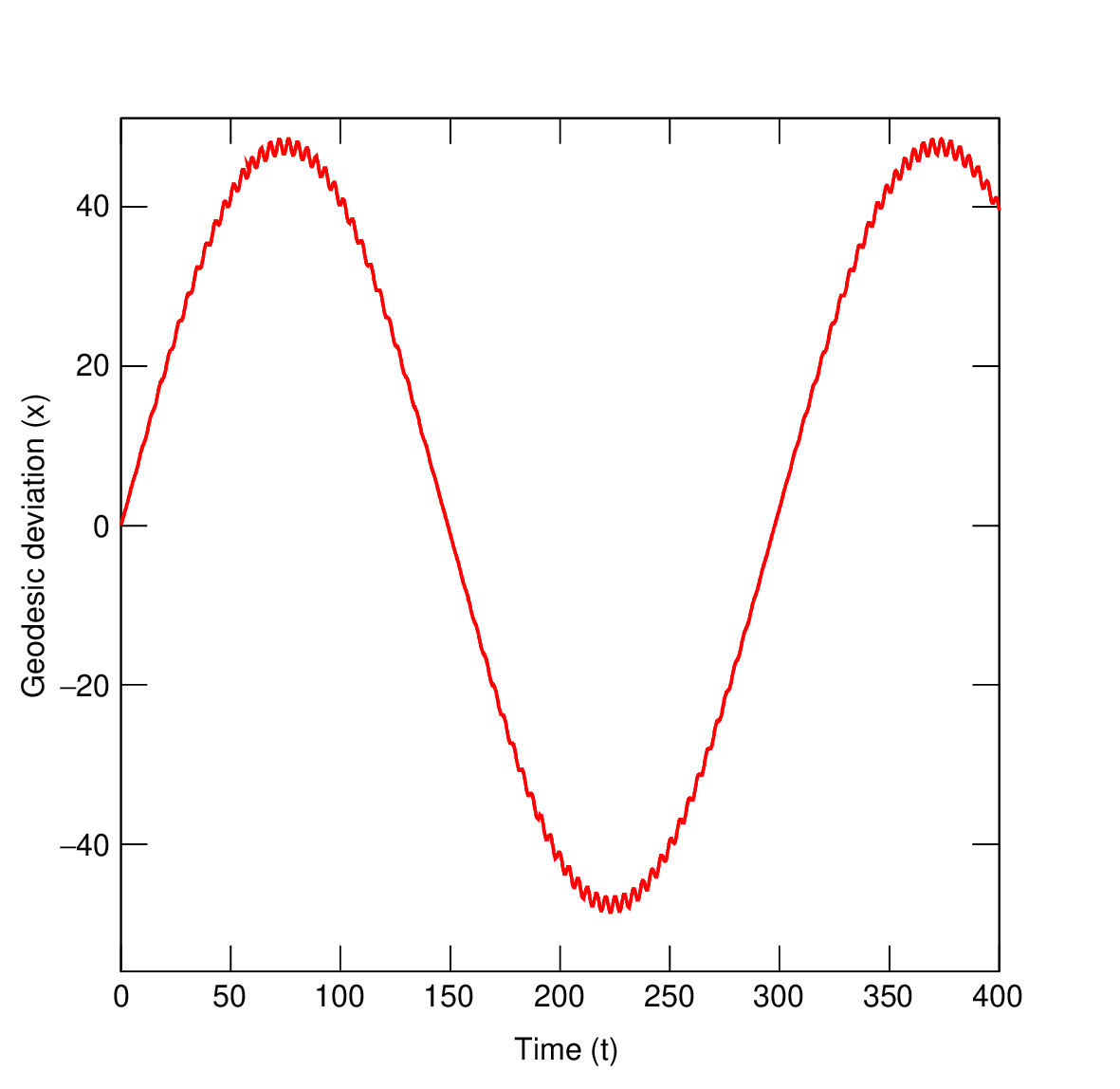}}
\centerline{
\includegraphics[scale=0.32]{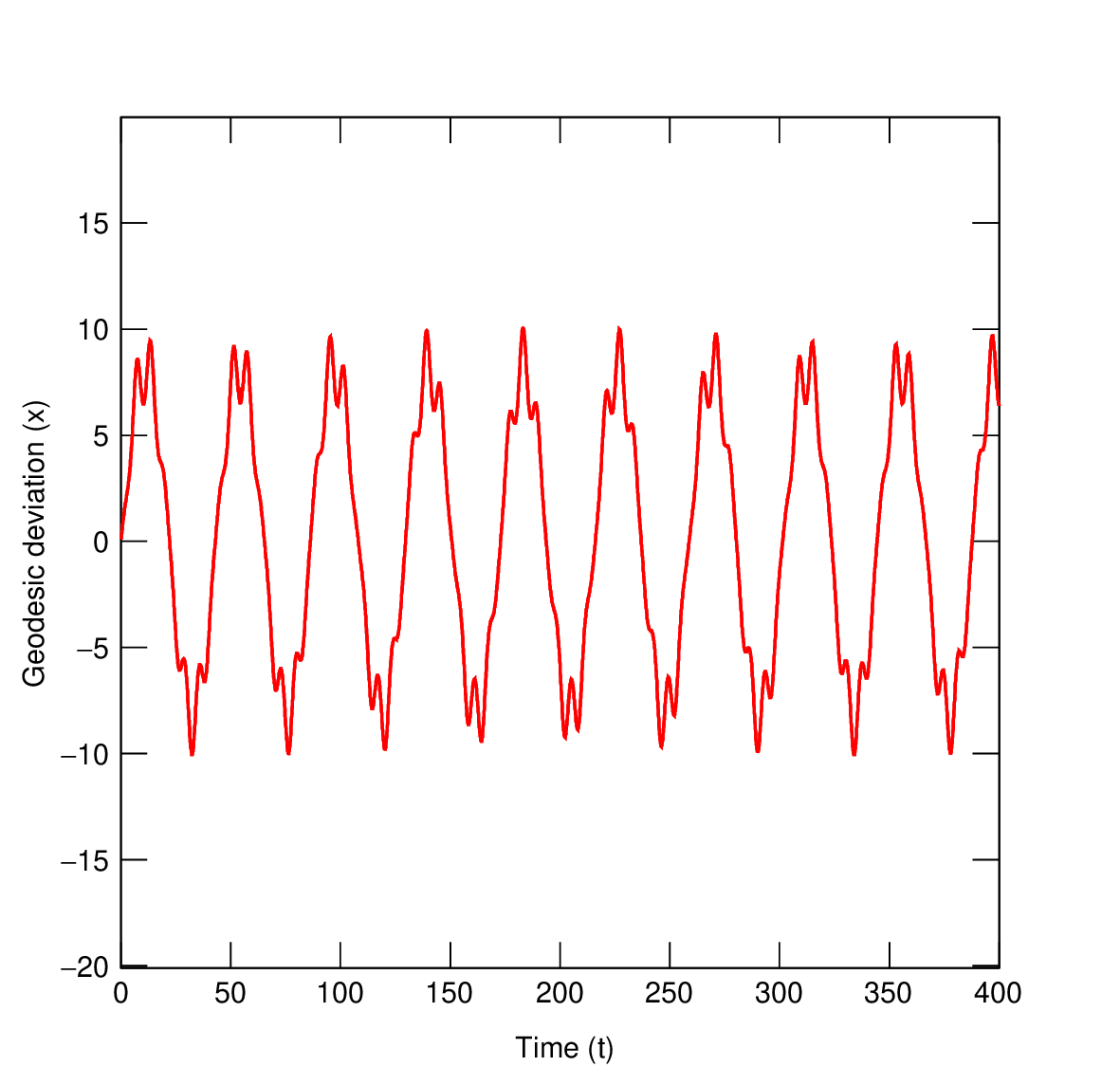}}
\caption{Geodesic deviation along $x$ for a fixed $z$. The top plot is for
$\omega = 1.5$, $\phi_0  = 0.02$ and $z= 1$, and the bottom is for 
$\omega = 1$, $ \phi_0  = 0.2$ and $z = 1$. All parameters are in arbitrary units.}
\label{fig2}
\end{figure}

The de Sitter spacetime can be expressed as a hypersurface in the host pseudo-Euclidean space with
metric $\eta_{AB}=(+1, -1, -1, -1, -1)$ with points in Cartesian coordinates $\chi^A$ satisfying
$\eta_{AB}\chi^A\chi^B= -\, l^2$, here $l$ is the de Sitter length parameter. The four-dimensional 
stereographic coordinates ${x^\mu}$ are  obtained through a stereographic projection from 
the de Sitter hypersurface into a target Minkowski spacetime \cite{geo00}. 
Taking into account the local transitivity properties of spacetime, we 
consider 
a family of modified de Sitter geodesics that are able to connect any two
points in spacetime and they are satisfied by the equation \cite{geo01, geo02}
\begin{equation} \label{geodesic}
\dfrac{dU_\rho}{ds} - \Lambda^\gamma _{\rho\sigma} U_\gamma u^\sigma = 0,
\end{equation}
where $s$ is an affine parameter, $ u^\sigma$ is the usual $4$-velocity of 
particles and $\Lambda^\gamma _{\rho\sigma}$ is the Christoffel connection. 
$U^\mu = \xi^\mu_\alpha u^\alpha$ is the anholonomic $4$-velocity, which is 
related with the usual $4$-velocity of particles via the Killing vector 
$\xi^\mu_\alpha$. The geodesic deviation equation obtained from the 
Eq. (\ref{geodesic}) can be written as
\begin{equation}\label{geo_deviation}
\dfrac{D^2 V^\mu}{Ds^2}\delta\lambda = R^\mu_{\nu\sigma\rho} U^\nu u^\sigma \eta^\sigma + 
\dfrac{D}{Ds}\left[ \left( \eta^\alpha u^\beta - \eta^\beta u^\alpha\right) \triangledown_\beta \xi^\mu_\alpha\right],
\end{equation}
where $\dfrac{D}{Ds} = u^\sigma \triangledown_\sigma$ represents the covariant 
derivative with respect to $s$, $V^\mu=\xi^\mu_\alpha v^\alpha$, $\eta^\mu = \dfrac{\partial x^\mu}{\partial \lambda}\delta \lambda = v^\mu \delta \lambda$ and $\lambda$ is some parameter such that for each $\lambda =$ constant, $x^\mu = x^\mu(s,\lambda)$ will represent the equation of a geodesic in Eq. (\ref{geodesic}). This equation shows that geodesic 
deviation in de Sitter spacetime contains translational (first term on RHS) as 
well as rotational (second term on RHS) terms \cite{geo03}. Since, the second 
term on RHS is model independent as it does not contain Riemannian tensor term,
the rotational effect is a generic property of the spacetime only. That is why, 
in Newtonian limit with $l\rightarrow \infty$ we can safely neglect the second 
term on RHS and thus can have an approximated and simplified form of geodesic 
deviation equation with only first term on the RHS. Now to see the geodesic 
deviation in our model, we first substitute the solution (\ref{eq36})
in the Eq. (\ref{eq34}), which gives,
\begin{equation}\label{eq38}
R_{\mu\nu} = -\, \dfrac{1}{4}\, \bar{g}_{\mu\nu}\,\delta\phi - p_\mu p_\nu. 
\end{equation}
For the wave traveling along the $z$ direction, the nonzero components of 
$R_{\mu\nu}$ are $R_{tt}$, $R_{tz}$ and $R_{zz}$. Now to linear order we 
may write,
$$ R_{\mu\nu\alpha\beta} \approx \dfrac{1}{2} ( h_{\nu\alpha,\mu\beta}
+h_{\mu\beta,\nu\alpha} - h_{\mu\alpha,\nu\beta} - h_{\nu\beta,\mu\alpha}).$$
This can be further simplified as
\begin{equation} \label{eq39}
 R_{itjt} = -\, \dfrac{1}{2}(h_{ij,tt} + h_{tt, ij}).
\end{equation}
Thus, using only the scalar part of the Eq. (\ref{eq37}) we can write the 
simplified and approximated geodesic deviation equation from the Eq. (\ref{geo_deviation})
in the limit $l\rightarrow \infty$ as
\begin{equation}\label{eq40}
\ddot{x}^i = \dfrac{1}{2}\,\Big(\delta^i_j \delta\ddot{\phi} - \delta\phi_{\,,j}^{\,,i}\Big)x^j,
\end{equation}
which gives, 
\begin{align}\label{eq41}
 \ddot{x}& = \frac{1}{2}\,\delta \ddot{\phi} x,\\  
 \label{eq42}
 \ddot{y}& = \frac{1}{2}\,\delta \ddot{\phi} y,\\  
 \ddot{z}& = 0.
\end{align}
Thus the geodesic equations show that there is no longitudinal component of 
the GW polarization for the pure $R^2$ model. So, for this model besides the 
tensor modes there exists only a scalar mode which is massless and pure in 
nature, known as the breathing mode. Again, from the Eq. (\ref{eq37}) 
considering only the scalar part we may write,
\begin{equation}\label{eq43}
\delta \ddot{\phi}=-\,\omega^{2}\delta\phi.
\end{equation}
Using this equation in the Eq. (\ref{eq41}) we have,
\begin{equation}\label{eq44}
\ddot{x}+\frac{1}{2}\,\omega^{2}\delta\phi\, x = 0,
\end{equation}
which for a GW propagating along $z$ axis takes the form:
\begin{equation}\label{eq45}
\ddot{x}+\frac{1}{2}\,\phi_{0}\,\omega^{2}\cos\omega(t-z) x = 0.
\end{equation}
The solution of this Eq. (\ref{eq45}) gives the time variation of 
deviation of $x$ at some fixed $z$. This equation is a special form of the 
well known Mathieu's equation \cite{canosa_numerical_1971, gheorghe_charged_0000} and the solution is graphically shown in Fig.\ref{fig2} for two different 
set of parameters. It is seen that the time period of geodesic deviation 
depends highly on $\omega$. The results are identical along the
$y$ direction also as clear from the Eq. (\ref{eq42}).

%\marginpar{Fig. 02 about here}

\section*{6. Polarization Modes With Newman-Penrose Formalism}

The Newman-Penrose (NP) formalism \cite{newman_approach_1962} can be used to 
find out the different polarization modes of GWs in a model. However, one major
drawback of NP formalism is that it is only applicable to null waves. In $f(R)$
theory metric formalism, usually GWs have massive longitudinal mode of 
polarization due to which the NP formalism fails and shows deviated results 
\cite{liang_polarizations_2017}. But in the particular case of pure $R^2$ model,
as we have already seen that the scalar field is massless and thus this allows 
us to use the NP formalism in the study of polarization modes of GWs in the 
theory. In NP formalism a GW is described with the help of six amplitudes
$\lbrace \psi_2, \psi_3, \psi_4, \phi_{22} \rbrace$ representing six
polarization modes in a particular coordinate system or frame 
\cite{newman_approach_1962}. All these amplitudes are defined as 
\cite{eardley_gravitational-wave_1973}:
\begin{align}
\notag
\psi_2 &= -\,\dfrac{1}{6} R_{ztzt},\\\notag
\psi_3 &= -\,\dfrac{1}{2}R_{xtzt}+\dfrac{1}{2}i R_{ytzt},\\\notag
\psi_4 &= -\,R_{xtxt} + R_{ytyt}+ 2i R_{xtyt},\\\notag
\phi_{22} &= -\, R_{xtxt} - R_{ytyt}.
\end{align}
Each of the complex amplitudes $\psi_3$ and $ \psi_4 $ are actually equivalent 
to two real amplitudes \cite{eardley_gravitational-wave_1973}.
In Brans-Dicke theory, the massless scalar field appears as the breathing mode 
showing $ \phi_{22} = - \,R_{xtxt} -\, R_{ytyt}\neq 0.$ For our model these
amplitudes are found as 
\begin{align} \notag
\psi_2& = 0,\\\notag
\psi_3 &= 0, \\\notag
\psi_4 &= \ddot{\bar{h}}_{xx}-\ddot{\bar{h}}_{yy}+ i\,(2\,\ddot{\bar{h}}_{xy}),\\\notag
\phi_{22} &= -\,2\,\delta\ddot{\phi}. 
\end{align}
These results show that our model, i.e. pure $R^2$ model gives three modes
of massless polarization: two for tensor modes and one for the breathing mode as
found earlier, and according to Lorentz - invariant $E(2)$ classification of 
plane waves, the model results GWs of class $N_3$ and all modes are independent 
of the observer \cite{eardley_gravitational-wave_1973}. It should be mentioned
that a similar study was done in Ref. \cite{kausar_gravitational_2016} for the same toy model in Palatini formalism. 
In this case, using the NP formalism one can obtain,
\begin{align} \notag
\psi_2& = 0,\\\notag
\psi_3 &= 0, \\\notag
\psi_4 &= \ddot{\bar{h}}_{xx}-\ddot{\bar{h}}_{yy}+ i\,(2\,\ddot{\bar{h}}_{xy}),\\\notag
\phi_{22} &= 0. 
\end{align}
This shows that in Palatini formalism, the model $f(R)=\alpha R^2$ gives only tensor modes of polarization of GWs and there is no scalar mode of polarization or scalar degrees of freedom. 

\section*{7. Pulsar Timing Arrays and Correlation of Polarization Modes}
Pulsar timing arrays (PTAs) can be used to detect the polarization modes of
GWs. A good number of works have been going on in this field and PTAs are found to be effective
in the detection of extra polarization modes to be present in GWs. Hence they can 
be used as a tool to test modified theories of gravity \cite{lee_pulsar_2008, mingarelli_characterizing_2013, lee_detecting_2010}.

\begin{figure}[hbt]
\centerline{
\includegraphics[scale=0.35]{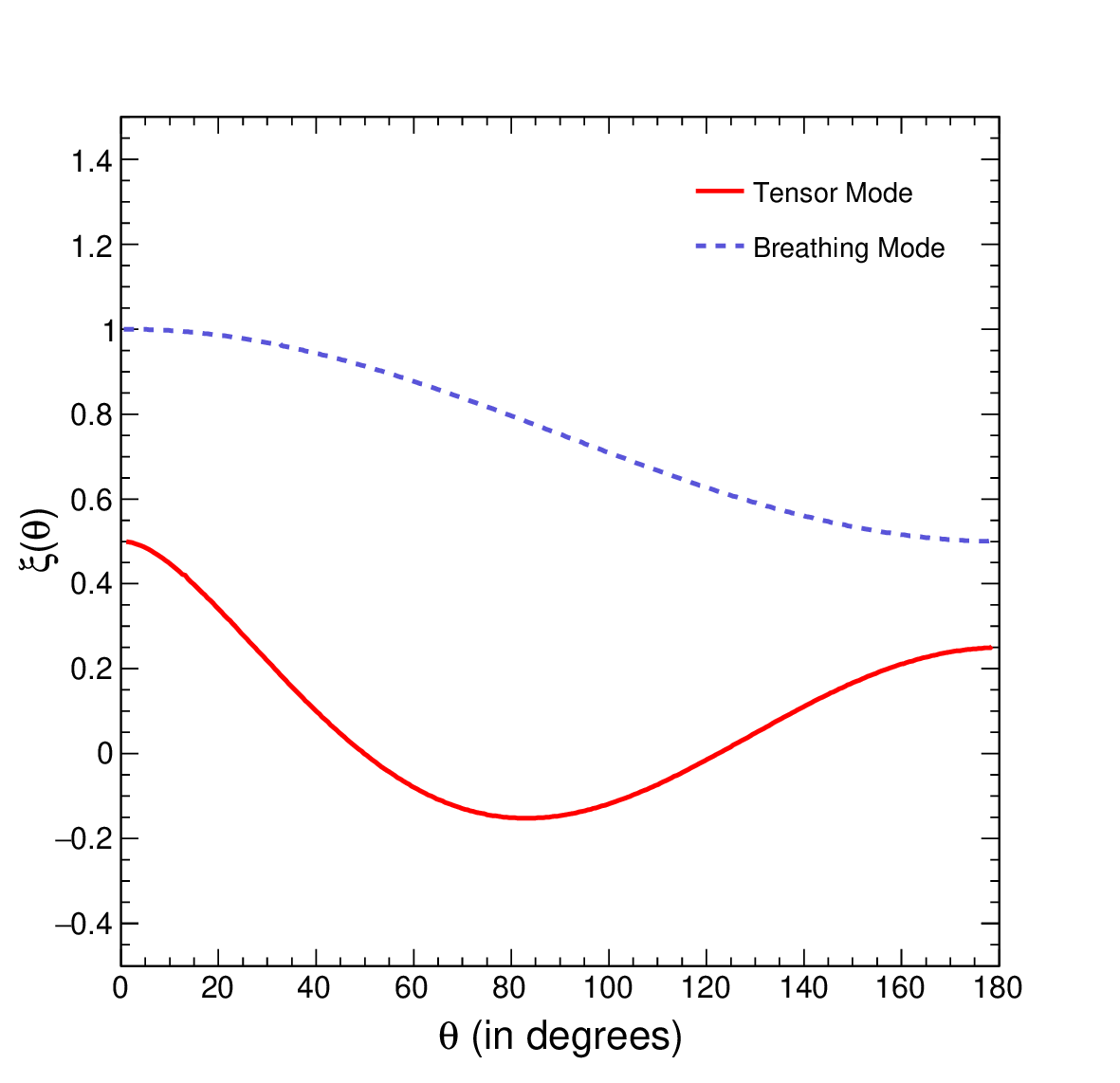}}
\caption{Variation of correlation coefficient $\xi(\theta)$ of different 
polarization modes with respect to angular separation between two pulsars.}
\label{fig3}
\end{figure}

Presence of GWs disturb the null geodesic of the signals from pulsars. Due to
this reason the time of arrival of pulsar signal changes. So, by tracing the changes in 
time of arrival of the radio signals from the pulsars, it is possible to detect 
the GWs. For the mathematical treatment of the PTA procedure let us consider a 
PTA detecting radio signals in the regime of GWs. In case of GWs from the pure 
$R^2$ model, the information about the source is carried by the three 
polarization amplitudes, viz., $h_+(t)$, $h_\times(t)$ and $h_\phi(t)$, where 
the first two stands for the Einstein or tensor modes and the third one for 
scalar (breathing) mode of polarization. Thus the GW signal from a source for 
the pure $R^2$ model can be wriiten as
\begin{equation}\label{eq46}
h_{ij}(t,\hat{\Omega}) = e^+_{ij}(\hat{\Omega})h_+(t,\hat{\Omega}) + e^\times_{ij}(\hat{\Omega})h_\times(t,\hat{\Omega}) + e^\phi_{ij}h_\phi(t,\hat{\Omega}).
\end{equation}
%\begin{equation}
%h_{ij}(t,\hat{\Omega}) = e^+_{ij}(\hat{\Omega})h_+(t,\hat{\Omega}) + e^\times_{ij}(\hat{\Omega})h_\times(t,\hat{\Omega}) +e^\phi_{ij} h_\phi(t,\hat{\Omega}).
%\end{equation}
%Here $e^+_{ij}$, $e^\times_{ij}$ and $e^\phi_{ij}$ are the polarization tensors which are uniquely 
Here $e^+_{ij}$ and  $e^\times_{ij}$ are the polarization tensors as
given by,
\begin{eqnarray}\nonumber
e^+_{ij}(\hat{\Omega}) = \hat{m}_i \hat{m}_j - \hat{n}_i \hat{n}_j,\\\nonumber
e^\times_{ij}(\hat{\Omega}) = \hat{m}_i \hat{n}_j + \hat{n}_i\hat{m}_j.
\end{eqnarray}
Hence, these polarization tensors are uniquely defined once the unit vectors 
$\hat{m}$ and $\hat{n}$ used to describe the principal axes of wave are 
specified. $\hat{\Omega}$ is the direction of 
GW propagation given by $\hat{m} \times \hat{n}$. As per our previous 
convention, we set $\hat{\Omega} = \hat{z}$, and consider a null vector
 $S^\mu$ that points to the Solar system barycenter from the pulsar in 
Minkowski spacetime. In perturbed spacetime vector $S^\mu$ will change to 
another vector $\sigma^\mu$ as given by,
\begin{equation}\label{eq47} 
\sigma^\mu = S^\mu - \dfrac{1}{2}\eta^{\mu\nu}h_{\mu\nu}S^\nu + \delta \phi\, \eta^{\mu\nu}b_{\mu\nu} S^\nu, 
 \end{equation}
where the second part of this equation is due to tensor modes of perturbation
and the third part is due to the scalar mode of perturbation in spacetime with
$b_{\mu\nu}$ is a unit breathing mode matrix having two non-zero unit 
components 
$b_{11}$ and $b_{22}$, obtained by the application of transverse condition to 
this mode of GWs. It is to be noted that for this mode we can not apply the 
traceless 
condition as the application of this condition to this mode will retain only
the tensor modes or GR modes \cite{kausar_gravitational_2016}. Now, If
we define $S^\mu$ as $S^\mu = \nu(1,-\alpha,-\beta,-\gamma)$, where $\alpha,\beta$ and $\gamma$ are direction cosines, and $\nu$ is the frequency of the radio
pulses from the source, then from Eq. (\ref{eq47}), we may write,
\begin{align}\notag 
\sigma^t& = \nu\,\\\notag
\sigma^x &= -\,\nu\,\big[\alpha\,\big(1-\dfrac{1}{2}h_+ - \delta\phi\big)-\dfrac{\beta}{2}h_\times\big],\\\notag
\sigma^y& = -\,\nu\,\big[\beta\big(1+\dfrac{1}{2}h_+ + \delta\phi\big)-\dfrac{\alpha}{2}h_\times\big],\\\notag
\mbox{and}\;\;
\sigma^z &= -\,\nu\,\gamma.
\end{align}
The radio pulses from the pulsar follow a null geodesic through spacetime. The 
geodesic equation of the pulses with the affine parameter $\lambda$ is
\begin{equation}\label{eq48}
\dfrac{d\nu}{d\lambda} = -\,\nu^2\alpha\,\beta\,\dot{h}_\times + \dfrac{1}{2}\,\nu^2 \big[\beta^2(\dot{h}_+ - \delta\dot{\phi}) - 2 \alpha^2 (\dot{h}_+ + \delta\dot{\phi}) \big].
\end{equation}
Using this equation in the geodesic deviation equation we can have the 
frequency shift at Solar system barycenter as given by,
\begin{align}\label{eq49}
z(t, \hat{\Omega}) &=\; \mid \dfrac{\nu(t)- \nu_0}{\nu_0} \mid \\\notag
&= \dfrac{1}{2(1+\gamma)}\Big[\alpha^2 (\Delta h_+ + \Delta \delta\phi)
 - \beta^2(\Delta h_+ - \Delta \delta \phi)\Big] \\ \notag
&  +\dfrac{\alpha \beta}{(1+\gamma)}
 \Delta h_\times,
\end{align}
where $\nu(t)$ is the frequency observed at Solar system barycenter. Thus it is seen that the presence of GWs will give rise to a frequency shift and these shifts can be observed with the help of PTAs. It is also clear that the presence of 
massless breathing mode results in a contribution of a shift in the frequency 
besides a contribution from the usual tensor (GR) modes. Following
\cite{lee_pulsar_2008}, for stochastic GW background with normalized frequency
$f$, the correlation function for GR modes and breathing mode of polarization 
are calculated. This function for tensor modes is found as
\begin{equation}\label{eq50}
C^{+,\times}(\theta) = \xi^{GR}(\theta)\int_0^\infty\dfrac{|h_c^{+,\times}|^2}{24\pi^2 f^3}\,df,
\end{equation}
where $$\xi^{GR}(\theta)=\dfrac{3\,(1-\cos\theta)}{4}\log\,(\dfrac{1-\cos\theta}{2}) + \dfrac{1}{2} - \dfrac{1-\cos\theta}{8} + \dfrac{\delta(\theta)}{2},$$
and $\theta$ is the angular separation between two pulsars. For the scalar 
modes it is
\begin{equation}\label{eq51}
C^{b}(\theta) = \xi^{b}(\theta)\int_0^\infty\dfrac{|h_c^{b}|^2}{12\pi^2 f^3}\,df,
\end{equation}
where $$\xi^{b}(\theta) = \dfrac{1}{8}\big[\cos\theta + 3 + 4\,\delta(\theta)\big].$$
The correlation coefficients $\xi^{GR}$ and $\xi^{b}$ as a function of 
$\theta$ are plotted in the Fig.\ref{fig3}. It is seen that provided the cases 
in \cite{lee_pulsar_2008} hold good, PTAs can effectively distinguish between 
the breathing mode and tensor modes present in GWs. It needs to mention that 
the correlation versus angular separation curve for the tensor modes of 
polarizations was first obtained by Hellings and Downs in 1983 and hence it is
usually known as Hellings-Downs (HD) curve \cite{hellings_upper_1983, jenet_understanding_2014}.

%\marginpar{Fig. 03 about here}

The calculations of correlation functions are model independent, but for different 
polarization modes there are different correlation functions \cite{lee_detecting_2010}. 
Since our toy model $f(R)=\alpha R^2$ gives only tensor plus, cross and massless breathing modes of 
polarization of GWs, we have included the correlation functions for tensor modes and massless 
breathing mode only. This provides an experimental way to test the viability of the 
toy model. 
%However, in other $f(R)$ gravity models (in metric formalism), there are 
%four polarization modes viz. tensor plus, cross, massless breathing and massive 
However, in other $f(R)$ gravity models (in metric formalism), apart from the tensor plus and cross modes there exists a scalar polarization mode which is a mixed state of massive longitudinal mode and massless transverse breathing mode \cite{liang_polarizations_2017, kausar_gravitational_2016, Hou}. 
Therefore, in those models, the experimental viability can be checked by considering 
three correlation functions for tensor modes, massless transverse breathing mode and 
massive longitudinal mode. In the near future, with increased sensitivity and sufficient data, the 
experimentally obtained correlation functions from pulsar timing array data can hopefully 
help us to distinguish between different polarization modes which will provide us a way 
to check the viability of the toy model.

\section*{8. Conclusions} 

In this work, we have used the $f(R)$ gravity power law model to study the 
polarization modes of Gravitational Waves in de Sitter spacetime. We have found that the field equations in the de Sitter spacetime show the existence of a massive scalar mode with a mass term $m^2 = \dfrac{(2-n)\bar{R}}{(n-1)}.$ The mass term varies widely 
with the exponent term $n$ and the background curvature or the de Sitter 
curvature, and the mass of the scalar field becomes zero when background 
curvature is zero or $n=2$. Later using the stability condition for the theory 
in de Sitter spacetime we have seen that for a constant curvature $\bar{R}$ 
the theory has stable solutions for $n=2$. It has been observed that for this 
particular case of the model, the massive longitudinal mode of polarization 
vanishes. Thus, this is the only case in $f(R)$ theory in metric formalism 
treatment where the third scalar mode 
is a pure breathing mode and there is no massive longitudinal mode present. To validate this result we've studied the geodesic 
deviations for the pure $R^2$ model explicitly and later we used the NP formalism 
to confirm the validity of our result. The absence of the massive scalar field in 
this model allows us to use the NP formalism, which is a powerful tool to check 
the polarization contents of null GWs. The results from the analysis show 
absence of massive polarization mode of the GWs in the theory, which establish
that the polarization modes and the mass of the scalar field in $f(R)$ gravity 
are model dependent. Another important result obtained from this work is that 
the scalar field is independent of the background curvature in 
pure $R^2$ model of $f(R)$ gravity and evidently chameleonic behaviour is not 
observed here. The absence of massive longitudinal mode makes this pure 
$R^2$ model different from the other $f(R)$ gravity models and hence a more 
detail study in this model is required. This model passes the cosmological bounds
from GW170817 \cite{Nojiri-gw}.

In a recent study \cite{Katsuragawa}, it was reported that due to the screening mechanism of the atmosphere
it might be difficult to observe the scalar mode of polarization at ground based GWs 
detectors. However, this can be possible only if the model shows chameleonic behaviour
which makes the mass of the scalar field background curvature dependent. In case of 
power law model, Eq. \eqref{eq10} shows the background curvature dependency of the
additional scalar field associated with the theory. The screening effect of the
atmosphere due to chameleonic behaviour can be measured if it is possible to
detect the scalar modes of such a theory both at ground based detectors and space
based detector. But the pure $R^2$ model does not have any massive longitudinal
mode of polarization and the associated scalar field in this case is massless in
nature. Due to this reason, the atmospheric screening can't be seen in this
case. 

The results of this study are also supported by another very recent study \cite{capp}, in
which the authors studied the GWs in teleparallel gravity explicitly.
Their study shows that there are three degrees of freedom associated with
teleparallel gravity and the third one is a scalar degrees of freedom. There
are two scalar modes of polarizations viz. massive longitudinal mode and
massless breathing mode in a mixed state in the theory. The authors have also 
mentioned that according to dynamics $f(T, B) \equiv f(R)$. Thus $f(R)$ gravity
shares the same results for polarization modes with $f(T, B)$ gravity 
\cite{liang_polarizations_2017, kausar_gravitational_2016, capp}. In this
study, we have shown that for a particular case the mixing of longitudinal mode and
breathing mode might not be there and the massive longitudinal mode vanishes. Similar results
can be expected from $f(T, B)$ gravity also, where there would be only three polarization modes.

It is worth to mention that $f(R)$ gravity with extra degrees of freedom can affect the
cosmological dynamics \cite{capo2, capo3}. Existence of extra polarization modes of GWs
can impose significant effect on the stochastic cosmological background.
The amplitude of GWs generated from inflation
also depends on the choice of $f(R)$ gravity model used for the study \cite{capo2}.
In Ref. \cite{capo3}, using $f(R)$ gravity power law model, it was shown that the
amplitude evolution of the tensor mode of GWs depends on the cosmological background.
However, it will be very premature to  comment on the cosmological background due to variation of polarization modes of GWs. We believe a further study will shed more light on this.

The pure $R^2$ model can be extended by adding other terms to the action which 
can account for the missing part giving rise to mixed polarization states of 
GWs besides tensor modes of polarization \cite{liang_polarizations_2017, kausar_gravitational_2016}. Those extended models can be constrained with the 
experimental results obtained so far \cite{jana}. It is expected that future experiments 
can provide better constraints to the $f(R)$ gravity models using which the 
existence of extra polarization modes can be hopefully confirmed, which in turn
give options to test the reliability of $f(R)$ gravity in the modifications 
and extensions of GR. Extensions like nonminimal matter field coupling and 
other modified gravity theories can be included as the future aspects
of this type of works, which can be tested with future experiments on GWs.

\section*{Acknowledgments}
UDG is thankful to the Inter-University Centre for Astronomy and Astrophysics 
(IUCAA), Pune, India for hospitality during his visits as a Visiting Associate.

\end{document}